\documentclass[10pt,aps,nofootinbib,
amsmath,amssymb,twocolumn,
preprintnumbers, 
showpacs,
raggedbottom, 
superscriptaddress,
prd,
floatfix]{revtex4-2}

\usepackage{cases}
\usepackage{pgf,tikz}
\usepackage{graphicx}
\usepackage{booktabs}
\usepackage{array}

\usepackage{multirow}
\usepackage{slashed} 

\usepackage{hyperref}
\usepackage[nameinlink]{cleveref}
\crefdefaultlabelformat{#2\textup{#1}#3}
\creflabelformat{equation}{#2\textup{#1}#3}

\newcommand{\ud}{\mathrm{d}}

\newcommand{\lrp}[1]{\left(#1\right)} 
\newcommand{\lrb}[1]{\left[#1\right]} 
\newcommand{\lrc}[1]{\left\{#1\right\}} 

\newcommand{\GeV}{\text{GeV}}

\newcommand{\msol}{M_\odot}

\newcommand{\tov}{\text{TOV}}
\newcommand{\pqcd}{\text{pQCD}}

\newcommand{\ceft}{{\chi\text{EFT}}}

\newcommand{\eden}{\mathcal{E}}

\newcommand{\cmin}{C_{s,\mathrm{min}}}
\newcommand{\cmax}{C_{s, \mathrm{max}}}
\newcommand{\cmean}{C_{s, \mathrm{mean}}}

\newcommand{\maxcmin}{\mathrm{max}\lrc{C_{s,\mathrm{min}}}}

\newcommand{\maxcmean}{\mathrm{max}\lrc{C_{s,\mathrm{mean}}}}

\newcommand{\maxallcmin}{\mathbf{max}\lrc{C_{s,\mathrm{min}}}}

\newcommand{\dpmax}{\Delta P_\mathrm{max}}
\newcommand{\dpmin}{\Delta P_\mathrm{min}}
\newcommand{\dpmean}{\Delta P_\mathrm{mean}}

\newcommand{\onset}{\mathrm{onset}}

\newcommand{\minc}[1]{\min\lrc{#1}}

\newcommand{\UCB}{Department of Physics, University of California Berkeley, Berkeley, CA 94720}

\begin{document}

\title{Bounds on the minimum sound speed above neutron star densities}
\author{Dake Zhou}
\email{dkzhou@berkeley.edu}

\affiliation{Department of Physics, University of Washington, Seattle, WA 98195}
\affiliation{Institute for Nuclear Theory, University of Washington, Seattle, WA 98195}
\affiliation{\UCB}

\date{\today}

\begin{abstract}

We show that the existence of massive neutron stars and asymptotic freedom of QCD place robust upper bounds on the lowest sound speed of the ultra-dense matter unattainable in neutron stars. 
Centered on worst-case scenarios, 
our limits are the most conservative among physical equations of state in the density range $\sim 2-40 n_0$.
Discovery of $\gtrsim 2.6 M_\odot$ neutron stars, in combination with current multimessenger astrophysical constraints on the equation of state, would strongly support first-order phase transitions at high baryon densities.

\end{abstract}

\maketitle

\section{Introduction}

Recent astrophysical observations~\cite{Demorest:2010bx,Antoniadis:2013pzd,Romani:2021xmb,NANOGrav:2019jur,Fonseca:2021wxt,Watts:2016uzu,TheLIGOScientific:2017qsa,LIGOScientific:2017ync,Miller:2019cac,Riley:2019yda,Miller:2021qha,Riley:2021pdl,Salmi:2024aum,Dittmann:2024mbo} confirming neutron stars (NSs) with masses exceeding two solar masses and radii likely smaller than $\sim13$~km have imposed stringent constraints on the equation of state (EOS) of matter at baryon densities in the range $n_B \sim 2 - 10n_0 $, where $n_0 = 0.16~\mathrm{fm}^{-3}$ is the nuclear saturation density~\cite{LattimerPrakash:2010,Abbott:2018exr,Tews:2018aa,De:2018uhw,Radice:2017lry,Capano:2019eae,Essick:2020flb,Legred:2021hdx,Huth:2021bsp,Rutherford:2024srk}.
These findings point to a physically compelling and observationally grounded scenario: the EOS exhibits relatively low pressure around $n_0$, while becoming significantly stiffer at higher densities in order to stabilize massive NSs. As a result, the squared adiabatic speed of sound, $C_s = c_s^2 = \partial P/\partial\eden$, where $P$ and $\eden$ are the pressure and energy density, must increase markedly within the NS core. This rising trend in $C_s$ strongly constrains models that feature strong first-order phase transitions (FOPT), as such transitions typically induce a sharp reduction in the speed of sound -- a behavior now increasingly disfavored by multimessenger observations~\cite{Christian:2019qer,Huth:2021bsp,Gorda:2022lsk,Brandes:2023hma,Rutherford:2024srk}.

The aim of this letter is to address a compelling question that naturally arises in this context: {\it Is a first-order phase transition still viable at densities beyond those probed by neutron stars?} 
To explore this, we present an analysis that integrates precise NS mass measurements with a key feature of QCD: the speed of sound satisfies $C_s \leq 1/3$ at asymptotically high densities.
Previous studies~\cite{Bedaque:2014sqa,Tews:2018kmu,Drischler:2020fvz,Drischler:2021bup} have shown that the maximum of the sound speed squared, $\cmax$, likely exceeds $1/3$ within massive NSs in order to explain the existence of two-solar-mass pulsars~\cite{Demorest:2010bx,Antoniadis:2013pzd,NANOGrav:2019jur,Romani:2021xmb,Fonseca:2021wxt}. Since $C_s$ asymptotes to $1/3$ from below at very high densities, the requirement $\cmax > 1/3$ implies that $C_s$ must feature at least one minimum with $\cmin < 1/3$ at densities exceeding those found inside NSs, though its value remains undetermined.
The present letter fills this blank and places stringent upper limits on $\cmin$.

A robust upper bound on the minimum value of $C_s$ 
across the density range $n_B \sim 10-50~n_0$
can offer fresh insights into non-perturbative phenomena in dense quark matter. For example, it would enhance our understanding of the quark-hadron transition, which may manifest as a smooth crossover~\cite{Schafer:1998ef,Schafer:1999pb,Schafer:1999fe}, a first-order phase transition~\cite{Cherman:2018jir,Hirono:2018fjr,Cherman:2020hbe,Dumitrescu:2023hbe}, or a phase where hadronic and quark matter coexist over an extended range of densities~\cite{McLerran:2007qj,McLerran:2018hbz}.
When $\cmin$ is required to be low, such a constraint could inform the existence and characteristics of a FOPT at low temperatures, thereby guiding both theoretical frameworks and experimental efforts aimed at identifying the critical point within the QCD phase diagram~\cite{An:2021wof,Sorensen:2023zkk,Stephanov:2024xkn}.

Robust probes of the ultra-dense matter beyond the densities realized in NS are enabled by a global approach to the zero-temperature QCD phase diagram.
This strategy synthesizes constraints across a wide density regimes: nuclear theory at low densities, astrophysical observations at intermediate densities, and perturbative QCD (pQCD) 
~\cite{Freedman:1976xs,Freedman:1976ub,Vuorinen:2003fs,Kurkela:2009gj,Gorda:2018gpy,Gorda:2021kme,Gorda:2021znl,Gorda:2023mkk,Fernandez:2021jfr,Fernandez:2024ilg} at asymptotically high densities. 
Recent work has adopted a similar strategy to improve constraints on the NS EOS
~\cite{Annala:2017llu,Komoltsev:2021jzg,Gorda:2022jvk,Somasundaram:2022ztm,Zhou:2023zrm}. 
Based and extending on the model-independent framework in \cite{Zhou:2023zrm}, we derive the most conservative limit on $\cmin$ among physical possibilities in the density range $\sim 2-50n_0$.

Our main finding is that the maximum mass of NSs, commonly known as the Tolman–Oppenheimer–Volkoff (TOV) limit
 $M_\tov$~\cite{Tolman:1939jz,Oppenheimer:1939ne}, when combined with reliable theoretical constraints at both low and asymptotic densities, establishes robust bounds on $\cmin$ above NS densities.
Presently, the heaviest known pulsars are around $2M_\odot$~\cite{Demorest:2010bx,Antoniadis:2013pzd,NANOGrav:2019jur,Romani:2021xmb,Fonseca:2021wxt}.
While analyses of the post-merger evolution of GW170817~\cite{TheLIGOScientific:2017qsa} suggest $M_\tov \lesssim 2.2 -2.3~M_\odot$
~\cite{Margalit:2017dij,Shibata:2017xdx,Rezzolla:2017aly,Radice:2018xqa,Shibata:2019ctb}, given the systematic uncertainties inherent in the modeling of the remnant dynamics and EOS dependence, substantially larger values of $M_\tov$ cannot be robustly excluded.
In particular, the secondary component of GW190814 is inferred to be massive with mass $2.59^{+0.08}_{-0.09} M_\odot$ ~\cite{LIGOScientific:2020zkf}.
Assuming it is an NS, its implication for NS interiors has been explored previously in refs~\cite{Drischler:2020fvz,Godzieba:2020tjn,Fattoyev:2020cws}.
In light of ongoing and forthcoming gravitational wave observations, which are expected to expand the population of accurately measured NS masses~\cite{KAGRA:2013rdx,Kalogera:2021bya}, the correlation between $\cmin$ and $M_\tov$ we establish here offers a new pathway to study FOPTs and non-perturbative phenomena at densities beyond those encountered in NSs.

\section{The mean value bound}\label{sec:mvb}

\begin{figure}[h]
	\includegraphics[width=0.7\linewidth]{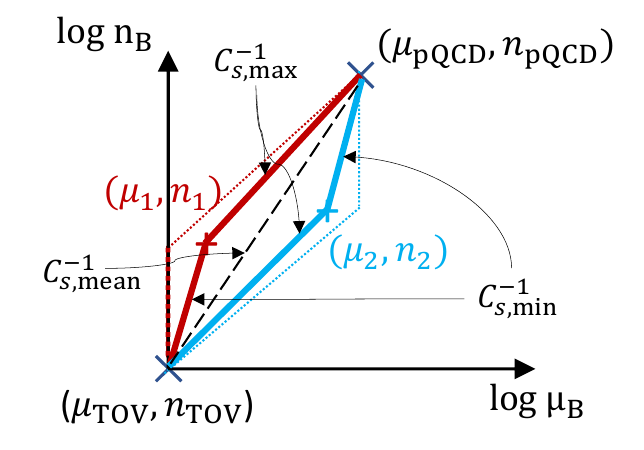}
	\caption{Schematics for the mean value bound \cref{eq:mvb} and the $\dpmax$ bound \cref{eq:dpmaxbound}.
    The slope of the dashed line is $\cmean^{-1}$ \cref{eq:csmean}.
	The maximally stiff and soft EOSs between the low- and high-density endpoints are shown in blue and red respectively.
	 Dotted lines show these limits when $\cmin=0$ and $\cmax=1$, and solid lines depict generalized bounds with $\cmin>0$ and $\cmax<1$.
	}\label{fig:demo}
\end{figure}

In the following, we employ thermodynamic principles together with the requirement of consistencies between the EOS at densities realized in NSs and its asymptotic behavior informed by pQCD. 
The first constraint exploits information about $n_B$ as a function of the baryon chemical potential $\mu_B$ at low ($L$) and high ($H$) densities.
At zero temperature, the slope of the function $n_B(\mu_B)$ is related to $C_s$ as 
\begin{equation}\label{eq:Cs}
C_s^{-1} = \frac{\ud \log n_B}{\ud \log \mu_B}\,.
\end{equation}
For any $n_B(\mu_B)$ relation that passes through $L$ and $H$, the mean value theorem from elementary calculus\footnote{The mean value theorem assumes $C_s$ is continuous and differentiable. A proof relaxing these assumptions is given in \cref{sec:proof}}
ensures that there exists a point within the interval $\mu_B \in [\mu_L, \mu_H]$ at which $C_s=\cmean$, where
\begin{equation}\label{eq:csmean0}
\cmean = \frac{\log\left(\mu_H / \mu_L\right)}{\log\left(n_H / n_L\right)}
\end{equation}
is the slope of the secant, the dashed black line in \cref{fig:demo}.
It follows that the minimum and maximum of $C_s$ between $L$ and $H$ 
denoted by $\cmin$ and $\cmax$ respectively, must bracket $\cmean$:
\begin{equation}\label{eq:mvb}
\cmin\leq\cmean\leq\cmax\,.
\end{equation}

In the absence of input from NS observations, theoretical predictions of $n_B(\mu_B)$ at low density from chiral effective field theory ($\ceft$)~\cite{Weinberg:1968de,Weinberg:1990rz,Weinberg:1991um,Weinberg:1992yk,Kaplan:1996xu,Kaplan:1998tg,Kaplan:1998we,Beane:2001bc}, and from $\pqcd$ at high density already provide an upper bound on $\cmin$.
Throughout this work, we take
$\mu_H=2.4$ GeV where pQCD 
appears to remain valid~\cite{Kurkela:2009gj,Gorda:2018gpy,Gorda:2022jvk,Fernandez:2021jfr}. We employ the next-to-next-leading order (N2LO) quark matter EOS~\cite{Freedman:1976ub,Freedman:1976xs,Kurkela:2016was} complemented by the $\mathcal{O}(\alpha_s^3\log^2\alpha_s)$ contribution at N3LO~\cite{Gorda:2021kme,Gorda:2021znl}.
To obtain an initial estimate of $\cmean$, we adopt the N3LO pure neutron matter calculation at $n_B=2n_0$~\cite{Drischler:2017wtt,Drischler:2020yad} for the low density point $L$.
The results are shown in the last column of \cref{tab:pqcd}, and are close to $1/3$.
On one hand, the first half of \cref{eq:mvb} suggests that the conformal limit $C_s<1/3$~\cite{Cherman:2009tw,Hohler:2009tv,Bedaque:2014sqa,Hoyos:2016cob,Ecker:2017fyh,Hoyos:2021uff,Jokela:2020piw,Gursoy:2017wzz,Anabalon:2017eri} is maximally saturated if not violated in dense QCD, an observation independent of any NS inputs.
On the other, the resulting bound on $\cmin\lesssim0.3$ is insignificant as a dip in $C_s$ below $1/3$ is expected behavior to approach pQCD predictions.

\begin{table}[htbp]
\setlength{\tabcolsep}{1em} 
{\renewcommand{\arraystretch}{1.25}
\centering
\begin{tabular}{ @{}c|c|c|c|c||c @{} }
\hline 
\multirow{2}{0pt}{$X$} & \multicolumn{2}{c|}{$P_\pqcd$} & \multicolumn{2}{c||}{$n_\pqcd$} & \multirow{2}{*}{$\cmean^{\ceft,\pqcd}$} \\ \cline{2-5}
    & GeV/fm$^3$ & $P_\mathrm{FD}$ & $n_0$ & $n_\mathrm{FD}$ &  \\
\hline 
$1$  & 1.24 &	0.306	& 30.5 & 0.723 &  $0.308_{-0.011}^{+0.014}$ \\
$2$  & 2.68 &	0.662	& 31.3 & 0.742 &  $0.305_{-0.011}^{+0.014}$\\
$4$  & 3.07 & 	0.757	& 33.5 & 0.793 &  $0.298_{-0.011}^{+0.014}$\\
\hline 
\end{tabular}
}
\caption{ 
The mean value between pQCD at $\mu_B=2.4$ GeV and $\ceft$ at $n_B=2n_0$.
$P_\pqcd$ and $n_\pqcd$ are shown as ratios to those of a non-interacting gas of quarks following the Fermi-Dirac (FD) distribution at $\mu_B=2.4$ GeV,
and their uncertainties are estimated by varying the renormalization scale $X=\bar{\Lambda}/(\mu_B/3)$.
Quoted errors on $\cmean^{\ceft,\pqcd}$ stem from $2\sigma$ $\ceft$ uncertainties.
}
\label{tab:pqcd}
\end{table}

Further insights into $\cmin$ 
can be gleaned by incorporating astrophysical constraints on the EOS at densities ranging from $\sim n_0$
to the highest value realized inside NSs.
This highest density is reached at the center of the heaviest NSs known as the TOV limit~\cite{Tolman:1939jz,Oppenheimer:1939ne}.
Applying the mean-value bound \cref{eq:mvb} between this TOV density and pQCD regime therefore yields a robust constraint on $\cmin$ of the ultra-dense matter, where the mean value is
\begin{equation}\label{eq:csmean}
\cmean = \frac{\log\left(\mu_\pqcd / \mu_\tov\right)}{\log\left(n_\pqcd / n_\tov\right)}.
\end{equation}
Above, $n_\tov$ and $\mu_\tov$ are the baryon density and chemical potential at the TOV point, respectively.

\begin{figure}
\centering
\includegraphics[width=0.98\linewidth]{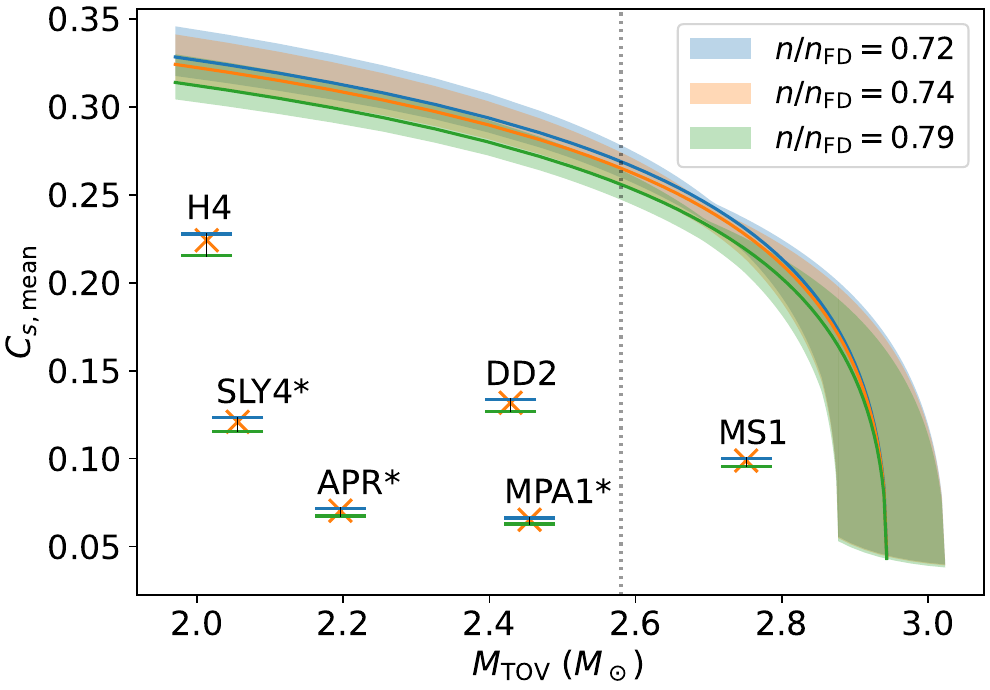}
\caption{Upper bounds on the mean value \cref{eq:csmean} maximized across NS inner cores. 
The bands represent uncertainties in the low-density EOS model adopted up to $n_c=2n_0$, and the colors reflect pQCD uncertainties  (see \cref{tab:pqcd}).
}
\label{fig:cmean}
\end{figure}

\Cref{fig:cmean} shows the mean value for a few NS EOSs widely adopted in the literature\footnote{Some models violate causality inside NSs and once that happens, we switch to the causal EOS defined by $C_s=1$. 
These are marked by asterisks in their labels.}. 
For instance, the APR* model~\cite{Akmal:1998cf} predicts $M_\tov^\mathrm{APR}=2.2~\msol$, $\mu_\tov^\mathrm{APR}=2.2~\GeV$ and $n_\tov^\mathrm{APR}=7.1~n_0$. 
By the mean-value bound \cref{eq:mvb}, the minimum squared sound speed cannot exceed $\cmean^\mathrm{APR}=0.070^{+0.002}_{-0.003}$, 
a strong indication of FOPT above $\simeq 7n_0$. 
Similar constraints obtain for other NS models, though the strengths 
vary across EOSs.
Quantifying uncertainties associated with modeling the NS inner cores is thus crucial to derive meaningful limits.
Here, we report a model-independent approach that focuses on the worst-case scenarios compatible with projected NS observations.
When future discoveries of heavier NSs demand $\cmin=0$ according to even the weakest bound among all physical possibilities in the NS core, they become a smoking gun signature of FOPT in dense QCD.

\section{The worst-case scenarios}\label{sec:ns}

To construct NSs, 
we follow the procedure detailed in refs~\cite{Zhou:2023zrm, Forbes:2019xaz} to describe the outer regions of NSs. 
While our low-density model is informed by $\ceft$ calculations~\cite{Drischler:2017wtt,Drischler:2020yad}, it remains sufficiently flexible to accommodate a broad class of scenarios consistent with experimental data~\cite{CREX:2022kgg,PREX:2021umo,Reed:2021nqk}, 
multimessenger astrophysical observations~\cite{Demorest:2010bx,Antoniadis:2013pzd,Romani:2021xmb,NANOGrav:2019jur,Fonseca:2021wxt,Watts:2016uzu,TheLIGOScientific:2017qsa,LIGOScientific:2017ync,Miller:2019cac,Riley:2019yda,Miller:2021qha,Riley:2021pdl,Salmi:2024aum,Dittmann:2024mbo}, and a wide range of dense matter models~\cite{Muther:1987xaa,Mueller:1996pm,Akmal:1998cf,Typel:2009sy,Douchin:2001sv,Lackey:2005tk}. As shown in \cref{fig:lowden} in \cref{sec:lowden}, the resulting pressure band exceeds the spread of $\ceft$ predictions and other constraints, with the exception of PREX, whose deviation is well established. Unless otherwise noted, we employ this prescription up to a baryon density of $n_c = 2n_0$, and defer a discussion of how projected constraints evolve with alternative choices of $n_c$ to the conclusion. 

Above $n_c$, finding the NS inner-core EOS that yields the weakest upper bound (i.e., the highest $\cmean$) consistent with given astrophysical observations belongs to the class of constrained optimization problem.
We find that $\maxcmean\equiv\max_\mathrm{NS}\cmean$, the upper bound on $\cmean$ across NS inner-core models, is reached by a construction known as the maximally stiff EOS in the literature~\cite{Rhoades:1974fn,Koranda:1996jm,Lattimer:2000nx,Drischler:2020fvz,Drischler:2021bup}.
It is specified by
\begin{equation} \label{eq:stiffNS}
    C_s(n_B)=
\begin{cases}
\cmax, \quad n_c\leq n_B\leq n_c+\Delta n_{\onset}\\
\cmin, \quad n_B> n_c+\Delta n_{\onset}
\end{cases}
\end{equation}
where $\Delta n_{\onset}$ is uniquely determined by the TOV limit $M_\tov$ upon solving the TOV equation~\cite{Tolman:1939,Oppenheimer:1939}. 

The resulting $\maxcmean$ are shown in \cref{fig:cmean} as colored bands.
They are largely insensitive to pQCD uncertainties (colors) because $n_\pqcd$ is tightly constrained by the asymptotic freedom of QCD (see \cref{eq:Cs}).
Variations in $n_\pqcd$ (in unit of $n_\mathrm{FD}$, the number density of free gas of quarks obeying Fermi-Dirac statistics) are shown to represent pQCD uncertainties, since $\cmean$ directly depends on it.
If higher $M_\tov$ is confirmed,
$\maxcmean$ decreases, thus strengthening the bound on $\cmin$. 
Meanwhile, uncertainties due to low-density EOS 
grows where a $\sim100\%$ error band on the pressure $P(n_B=2n_0)$ translates to $\sim 0.1M_\odot$ uncertainties on $M_\tov$ for the maximally stiff inner core~\cref{eq:stiffNS}.

\section{Bounds from pressure}\label{sec:dpb}

Stronger limits on $\cmin$ are possible by demanding that the pressure of weakly-coupled quark matter
$P_\pqcd$ can be reached from $P_\tov$, 
the central pressure of the most massive NS. 
Model-independent bounds on the pressure have been long known in the literature~\cite{Rhoades:1974fn,Koranda:1996jm,Lattimer:2000nx,Drischler:2020fvz,Drischler:2021bup,Komoltsev:2021jzg,Zhou:2023zrm}.
Here, we extend these previously-known results to account for the effect of $\cmin$, and show that they place stringent constraints on $\cmin$.

Above an arbitrary point in the EOS, 
the highest pressure at any density, or equivalently the lowest pressure at any chemical potential, is given by the maximally stiff EOS \cref{eq:stiffNS};
The opposite limit, namely the lowest (highest) pressure at any density (chemical potential), follows from the so-called maximally soft EOS.
For the purpose of constraining the EOS between TOV and pQCD points,
the maximally soft EOS takes the form
\begin{equation} \label{eq:soft}
    C_s(n_B)=
\begin{cases}
\cmin, \quad  n_\tov\leq n_B\leq n_1; \\
\cmax, \quad  n_1<n_B\leq n_\pqcd,
\end{cases}
\end{equation}
and is depicted in red in \cref{fig:demo}.
Above, $n_1$ is determined by the boundary condition $n_B(\mu_B=\mu_\pqcd)=n_\pqcd$, where the function $n_B(\mu_B)$ follows from integrating \cref{eq:Cs}.
Let $\alpha=\cmax^{-1}$ and $\beta=\cmin^{-1}$, one finds
\begin{equation}\label{eq:nmusoft}
n_B(\mu_B)=
\begin{cases}
n_\tov\lrp{\frac{\mu_B}{\mu_\tov}}^\beta, \quad &\mu_\tov \leq\mu_B\leq \mu_1\\
n_\pqcd\lrp{\frac{\mu_B}{\mu_\pqcd}}^\alpha, \quad &\mu_1 \leq\mu_B\leq \mu_\pqcd
\end{cases} 
\end{equation} 
where $\mu_1$ is the location at which the two segments intersect and $n_B(\mu_B=\mu_1)=n_1$. This intersection point is easily solved from \cref{eq:nmusoft}:
\begin{align}
	\mu_1&=\mu_\tov\lrp{\frac{\mu_\tov}{\mu_\pqcd}}^{\frac{\delta}{1-\delta}} \lrp{\frac{n_\pqcd}{n_\tov}}^{\frac{\cmin}{1-\delta}}\label{eq:mu1},\\
	n_1&=n_\tov \lrp{\frac{n_\pqcd}{n_\tov}}^{\frac{1}{1-\delta}} \lrp{\frac{\mu_\tov}{\mu_\pqcd}}^{\frac{\alpha}{1-\delta}}.\label{eq:n1}
\end{align}
Recall from \cref{eq:Cs} the slope of any curve in \cref{fig:demo} is $C_s^{-1}$, 
no EOS whose sound speed bounded by $\cmin\leq C_s(n_B)\leq\cmax$ may surpass the red curve.
Therefore, \cref{eq:nmusoft} is the highest possible $n_B(\mu_B)$ anywhere between $\mu_\tov$ and $\mu_\pqcd$, and decreases with increasing $\cmin$. 

To connect $\cmin$ with the pressure, we use the thermodynamic relation $\ud P/\ud \mu_B=n_B$ to obtain the pressure at $\mu_\pqcd$ relative to its value at $\mu_\tov$:
\begin{equation}\label{eq:dp}
\Delta P\equiv P(\mu_\pqcd)-P(\mu_\tov) =\int^{\mu_\pqcd}_{\mu_\tov} n_B(\mu_B)\ \ud \mu_B\,.
\end{equation}
Since \cref{eq:nmusoft} is the upper bound on the integrand $n_B(\mu_B)$, the maximally soft EOS leads to the highest possible $\Delta P$ given by
\begin{multline}\label{eq:dpmax}
	\Delta P_\mathrm{max}=\frac{n_\pqcd \mu_\pqcd}{\alpha+1}\lrb{1-\lrp{\frac{\mu_1}{\mu_\pqcd}}^{\alpha+1}}\\
	+\frac{n_\tov \mu_\tov}{\beta+1}\lrb{\lrp{\frac{\mu_1}{\mu_\tov}}^{\beta+1}-1}.
\end{multline}
where $\delta=\cmin/\cmax=\alpha/\beta$.

The lower bound $\Delta P_\mathrm{min}$ due to the maximally stiff construction (blue in \cref{fig:demo}) is obtained similarly. 
But since considerations based on $\dpmin$ 
are sensitive to the uncertain strength of non-perturbative effects in quark matter~\cite{Zhou:2023zrm}, 
we do not consider it in the main text and discuss it in \cref{sec:thermo}.

Since $\dpmax$ is the highest value compatible with thermodynamics, 
all physical possibilities must satisfy
\begin{multline}\label{eq:dpmaxbound}
 P_\pqcd-P_\tov \\
 \leq\Delta P_\mathrm{max}(\cmin,\ \cmax,\ n_\tov,\ \mu_\tov).
\end{multline}
If one assumes $\cmin=0$ and $\cmax=1$, \cref{eq:dpmaxbound} can rule out NS EOSs whose predicted TOV points are incompatible with \cref{eq:dpmaxbound}~\cite{Komoltsev:2021jzg,Gorda:2022jvk,Somasundaram:2022ztm,Zhou:2023zrm}.
Alternatively, one may stay agnostic about the ultra-dense matter and only assumes thermodynamic stability $\cmin\geq0$ and causality $\cmax\leq1$.
As such, \cref{eq:dpmaxbound} is a bound on $\cmin$ and $\cmax$ of the ultra-dense phase
informed by astrophysical observations. 
We adopt this second approach as multimessenger astronomy is expected to provide accurate NS measurements in the coming decade~\cite{Watts:2016uzu,KAGRA:2013rdx,Kalogera:2021bya}.

\Cref{eq:dpmaxbound} places an upper limit on $\cmin$ because its right-hand side $\dpmax(\cmin)$ is monotone decreasing, so that $\cmin$ may not exceed a certain threshold value $\maxcmin$.
To see this, notice that increasing $\cmin$ decreases the slope of the red $\cmin$ segment in \cref{fig:demo}, and this reduces the integrand $n_B(\mu_B)$ in \cref{eq:dp} 
hence lowers $\dpmax$.
The threshold value $\maxcmin$ saturates \cref{eq:dpmaxbound} and is given implicitly via the transcendental equation 
\begin{equation}\label{eq:maxCmin}
P_\pqcd-P_\tov = \Delta P_\mathrm{max}(\maxcmin).
\end{equation}

We note that 
\cref{eq:maxCmin} does not always admit a solution, either because \cref{eq:dpmaxbound} is violated even with $\cmin=0$, or when \cref{eq:dpmaxbound} is respected at $\cmin=\cmean$. 
In the first scenario, the NS EOS under consideration can be excluded~\cite{Komoltsev:2021jzg,Gorda:2022jvk,Somasundaram:2022ztm,Zhou:2023zrm}, or assuming that it is valid at low densities, FOPTs would already present below $n_\tov$ predicted by the given NS EOS~\cite{Zhou:2023zrm}; 
In the second case, $\maxcmin$ is to be derived from considerations based on $\dpmin$, see \cref{sec:thermo}. 
In either scenario, the mean value bound $\cmin\leq\cmean$ provides a conservative limit and stronger constrains exist.

\begin{figure}
\centering
\includegraphics[width=0.98\linewidth]{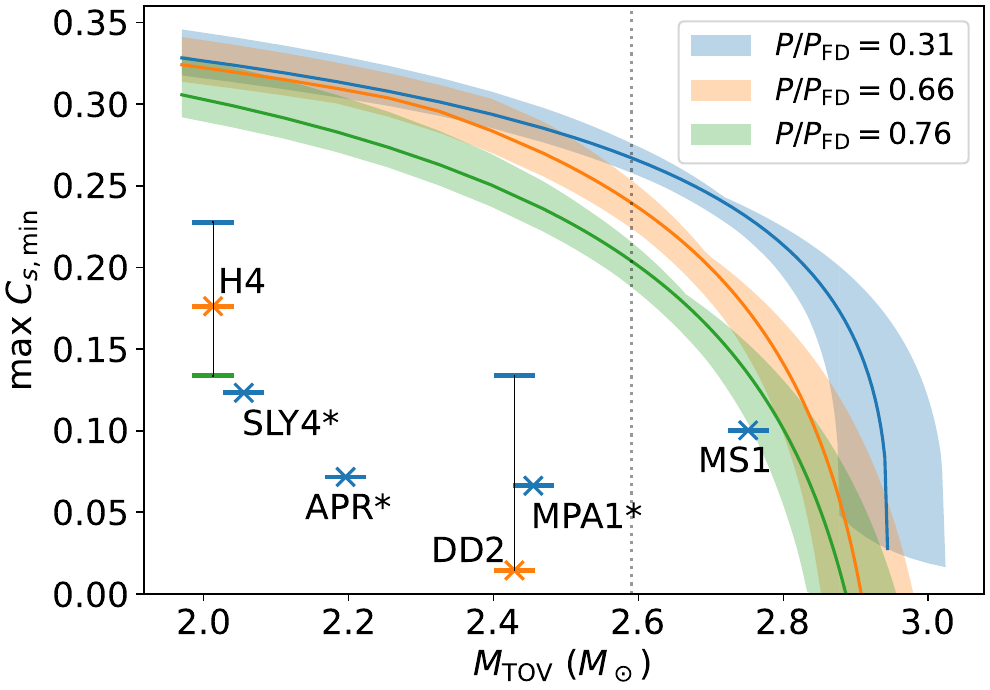}
\caption{Pressure-based bound $\maxcmin$ from \cref{eq:maxCmin}. The colored bands depict $\maxallcmin$, $\maxcmin$ maximized across NS inner-core EOSs similar to those in \cref{fig:cmean}.
}
\label{fig:cmin_cmax1}
\end{figure}

The upper bounds $\maxcmin$ for the set of named EOSs considered earlier are shown in \cref{fig:cmin_cmax1}.
Most of them are only compatible with rather low pressure in the quark matter with $P_\pqcd/P_\mathrm{FD}\lesssim0.5$.
For the worst-case scenarios, 
we find that $\maxallcmin\equiv\max_\mathrm{NS}\lrc{\maxcmin}$, the weakest bounds among NS inner-core models, are again produced by the maximally stiff EOS \cref{eq:stiffNS}. 
They are shown in \cref{fig:cmin_cmax1} as colored bands.
Compared to the bounds in ~\cref{fig:cmean}, the constraints here are stronger and can reach $\cmin=0$ for most pQCD predictions.
If NSs around $3\msol$ are found in future observations, sharp FOPT with $\cmin\simeq0$ at high densities would be strongly supported.


\begin{figure*} 
\includegraphics[width=\linewidth]{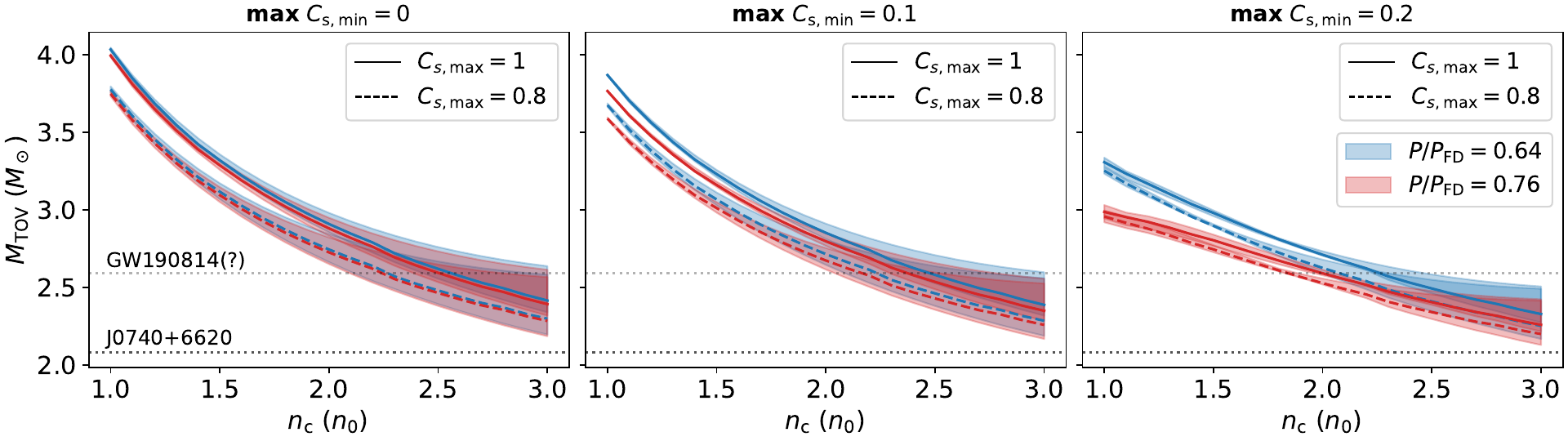}
\caption{
Identify QCD phase transitions 
via the existence of massive NSs.
Assuming current astrophysical constraints on the EOS up to $3n_0$, 
discovery of $\simeq2.6\msol$ NSs would confirm the presence of FOPT with $\cmin=0$ at high baryon densities.
}
\label{fig:cmin_nc}
\end{figure*}    

\section{discussion and conclusion}\label{sec:conclusion}

Our results can be simply understood by revisiting the mean values between $\ceft$ and $\pqcd$ densities, $\cmean^{\ceft,\pqcd}$ in \cref{tab:pqcd}. 
Since \cref{eq:csmean} is an ``averaged" slope of any $n_B(\mu_B)$ relation between $\ceft$ and pQCD densities,
 $C_s$ of an arbitrary EOS cannot diverge too much from $\cmean^{\ceft,\pqcd}$ ``on average", although sizable local deviations are permitted. 
Elevated $C_s$ over a wide density range required inside NSs to support a high $M_\tov$ thus must be compensated by segments of low $C_s$ above NS densities.

Future astrophysical observations in addition to NS mass measurements can also facilitate the search for FOPT at ultra-high densities.
In fact, incorporating current multimessenger constraints on the EOS in the density range $2-3n_0$  already produces stronger limits on $\maxcmin$, 
see \cref{fig:cmin_nc}.
For example, with $n_c=3n_0$, assuming the secondary component of GW190814 is indeed a NS, one finds that $\cmin$ cannot exceed $0.2$ regardless of pQCD or NS uncertainties, and that a sharp FOPT with $\cmin=0$ in dense QCD is strongly indicated. 
Additionally, upper limits on $\cmax$ such as the one proposed in \cite{Hippert:2024hum} also tighten the bound on $\cmin$,
although in most cases such improvements are degenerate with the current uncertainty associated with the NS EOS. 

A key ingredient underlies this work is the asymptotic freedom of QCD.
PQCD predicts a nearly conformal quark matter where $C_s^\pqcd$ remains close to $1/3$ and uncertainties well controlled even with naive extrapolations down to non-perturbative regimes~\cite{Vuorinen:2003fs,Kurkela:2009gj,Gorda:2021kme,Gorda:2021znl,Gorda:2023mkk,Fernandez:2021jfr}.
This leads to a tightly constrained $n_\pqcd$ but nevertheless poorly-determined $P_\pqcd$.
Refined understanding of weakly-coupled cold quark matter will improve the bounds on $\cmin$ and benefit the search for high-density FOPTs.
Furthermore, the near-conformal $C_s^\pqcd$ also renders seemingly uninformative values of $\maxcmin\simeq0.1-0.2$ intriguing, as they already deviate considerably from perturbative predictions. 
In this work, we have ignored non-perturbative effects in the dense quark matter, and this choice ensures the bounds are conservative.
Including superconducting pairing gaps~\cite{Alford:1997zt,Berges:1998rc,Carter:1998ji,Pisarski:1999bf,Son:1998uk,Alford:1998mk,Schafer:1999jg,Zhou:2023zrm,Kurkela:2024xfh} effectively increases $n_\pqcd$ in \cref{eq:csmean} and $P_\pqcd$ in \cref{eq:dpmaxbound}, thus reducing $\cmean$ and $\maxcmin$.
Strengthened constraints on $\cmin$ accounting for superconducting gaps and the $\dpmin$ bound
will be reported later.

\section*{Acknowledgment}

It is a pleasure to thank Sanjay Reddy and Michael Forbes for discussions and carefully reading a draft of the manuscript.
During the conception and completion of this work the author is supported 
by the Institute for Nuclear Theory Grant No. DE-FG02-00ER41132 from the Department of Energy,
and by NSF PFC 2020275 (Network for Neutrinos, Nuclear Astrophysics, and Symmetries (N3AS)).

\clearpage
\newpage
\appendix

\renewcommand\thefigure{\thesection.\arabic{figure}}    

\section{A proof of the mean value bound~\cref{eq:mvb}}\label{sec:proof}

The speed of sound squared $C_s(n)$ in thermodynamically stable matter may contain an arbitrary number of discontinuities or non-differentiable singularities, but must be an integrable function, whose integral yields the chemical potential. Therefore, by the fundamental theorem of calculus,
\begin{equation}
\begin{aligned}
\log\mu(n_H)-\log\mu(n_L)&=\int_{\log n_L}^{\log n_H} C_s(n)\ud \log n\\
& \geq\int_{\log n_L}^{\log n_H} \minc{C_s(n)}\ud\log n\\
&= (\log n_H-\log n_L)  \minc{C_s(n)},
\end{aligned}
\end{equation}
from which the first half of \cref{eq:mvb} follows:
\begin{equation}
  \cmin\equiv\minc{C_s(n)}\leq \frac{\log\mu(n_H)-\log\mu(n_L)}{\log n_H-\log n_L} \equiv \cmean.
\end{equation}
The proof for the second inequality in \cref{eq:mvb} follows similarly.

\section{the low-density EOS}\label{sec:lowden}
\setcounter{figure}{0}  

\begin{figure}[h]
\centering
\includegraphics[width=\linewidth]{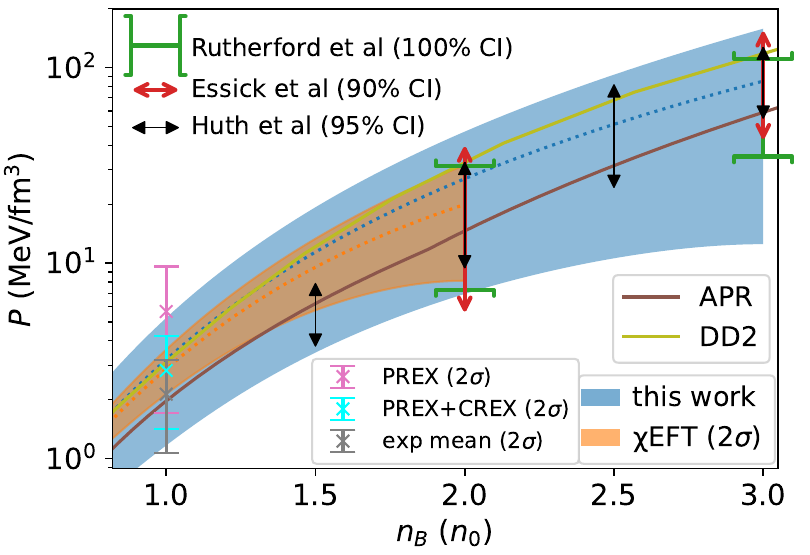}
\caption{
The range of low-density EOSs employed in this work (shaded blue) is consistent with and conservative than current constraints from astrophysics and nuclear physics.
The black arrows show the $95\%$ multimessenger posterior credible intervals (CIs) from ref~\cite{Huth:2021bsp}, and the red arrows indicate the $90\%$ CIs in ref \cite{Essick:2020flb} based on their ``completely agnostic'' model that does not utilize any nuclear theory inputs above crustal densities.
Shown in green are the $100\%$ posterior CIs from ref \cite{Rutherford:2024srk}, where the largest uncertainties across the 4 models reported therein (using chiral EFT up to $1.1n_0$ or $1.5n_0$, and adopting piecewise polytropic high-density models or piecewise speed of sound parameterizations) are chosen to obtain the most conservative uncertainty.
The orange band corresponds to the $2\sigma$ uncertainty of the N3LO $\ceft$ pure-neutron-matter calculation reported in ref \cite{Drischler:2020yad}.
The pink error bar represents the $2\sigma$ range of inferred $P_\mathrm{PNM}(n_B=n_0)=L/3 n_0$ from the PREX experiment~\cite{PREX:2021umo,Reed:2021nqk}, where $L$ is the slope of symmetry energy;
 The cyan and gray error bars are averaged CIs reported in refs \cite{Lattimer:2023rpe,Lattimer:2023xjm} combining PREX with CREX~\cite{CREX:2022kgg} plus additional experiments.  
}
\label{fig:lowden}
\end{figure}

In this section, we compare low-density EOSs adopted in this work with existing constraints from multimessenger observations of neutron stars, from nuclear theories,  and from nuclear experiments.
The range of $P(n_B)$ spanned by our model is shown in \cref{fig:lowden} between $n_0$ and $3n_0$.
\Cref{fig:cmean,fig:cmin_cmax1} in the main text is based on employing this class of model up to $n_c=2n_0$, and in \cref{fig:cmin_nc} $n_c$ is varied between $n_0$ and $3n_0$.
We do not attempt to assign probability distribution within this range, and the reported bands correspond to the extremal cases spanned by these possibilities.

Shown in orange is the recent N3LO chiral EFT calculation of pure neutron matter (PNM) using many-body perturbation theory with a $500$ MeV cutoff in \cite{Drischler:2017wtt}, along with the estimated $2\sigma$ truncation error \cite{Drischler:2020yad}. 
At $n_0$, our model predicts $P(n_0)$ in the range $1.1$ to $5.3$ MeV/fm$^{-3}$. Ignoring protons and electrons so that $P(n_0)\approx P_\mathrm{PNM}(n_0)$, this translates to a prediction for the slope of symmetry energy $L=3P(n_0)/n_0\in[21,99]$ MeV.
While in line with most experimental determinations of $L$, it is in mild tension with PREX measurements of the neutron skin thickness \cite{PREX:2021umo} that favor large $L$ \cite{Reed:2021nqk}.
The arrows show multimessenger posteriors on $P(n_B)$ at various densities reported in refs \cite{Essick:2020flb,Huth:2021bsp,Rutherford:2024srk}.
When several options are presented  in these analyses, we pick the most conservative choice (such as using the EOS model with the least amount of nuclear theory inputs) from each work.
The pressure band of our model are fully consistent with and even more conservative than these existing constraints.

Note that our model in fact spans a wider range in the lower half than shown in \cref{fig:lowden}. For instance, at $2n_0$ it predicts $P(2n_0)$ in the range $[3.8-43]$ MeV/fm$^{-3}$, fully enclosing the red arrow. 
The very soft EOSs in our model predicting low $P(2n_0)$ cannot be naively extended to $3n_0$ as they become unstable above $2n_0$.
But since they correspond to the lower boundaries in \cref{fig:cmin_nc}, the worst-case scenarios are not impacted by them.

\section{Details of $\Delta P_\mathrm{max}$ and $\Delta P_\mathrm{min}$}\label{sec:thermo}

Here we present the derivations and full expressions of $\Delta P_\mathrm{max}$ and $\Delta P_\mathrm{min}$ for generic $\cmin\geq0$ and $\cmax\leq1$. 
They are extensions to their previously-known counterparts that assume $\cmin=0$, see e.g., refs~\cite{Rhoades:1974fn,Koranda:1996jm,Lattimer:2000nx,Drischler:2020fvz,Drischler:2021bup,Komoltsev:2021jzg,Zhou:2023zrm}.

We begin by deriving $\Delta P_\mathrm{max}$ due to the maximally soft construction (red in \cref{fig:demo}).
The $n_B(\mu_B)$ relations for the constant $\cmin$ and $\cmax$ segments follow directly from \cref{eq:Cs}, and are given by
\begin{equation}
n_B(\mu_B)=
\begin{cases}
n_\tov\lrp{\frac{\mu_B}{\mu_\tov}}^\beta, \quad &\mu_\tov \leq\mu_B\leq \mu_1\\
n_\pqcd\lrp{\frac{\mu_B}{\mu_\pqcd}}^\alpha, \quad &\mu_1 \leq\mu_B\leq \mu_\pqcd
\end{cases}
\end{equation} 
where we have defined $\alpha=1/\cmax$, $\beta=1/\cmin$, and $\mu_1$ the baryon chemical potential at which the two segments intersect. This point of intersection is marked by the red ``+'' in \cref{fig:demo}, and its coordinate is found to be
\begin{align}
	\mu_1&=\mu_\tov\lrp{\frac{\mu_\tov}{\mu_\pqcd}}^{\frac{\delta}{1-\delta}} \lrp{\frac{n_\pqcd}{n_\tov}}^{\frac{1}{\beta-\alpha}},\\
	n_1&=n_\tov \lrp{\frac{n_\pqcd}{n_\tov}}^{\frac{1}{1-\delta}} \lrp{\frac{\mu_\tov}{\mu_\pqcd}}^{\frac{\alpha}{1-\delta}}.
\end{align}
Utilizing \cref{eq:dp},  we arrive at
\begin{multline}
\dpmax=\int^{\mu_1}_{\mu_\tov}\ud\mu\ n_\tov\lrp{\frac{\mu}{\mu_\tov}}^\beta
\\
+
\int^{\mu_\pqcd}_{\mu_1}\ud\mu\ n_\pqcd\lrp{\frac{\mu}{\mu_\pqcd}}^\alpha,
\end{multline}
which evaluates to \cref{eq:dpmax}.
This is the largest increment in pressure from $\mu_\tov$ to $\mu_\pqcd$.

For the maximally stiff EOS (blue in \cref{fig:demo}),
\begin{equation} \label{eq:stiff}
    C_s(n_B)=
\begin{cases}
\cmax, \quad  n_\tov\leq n_B\leq n_2; \\
\cmin, \quad  n_2<n_B\leq n_\pqcd.
\end{cases}
\end{equation}
since it only differs from $\Delta P_\mathrm{max}$ in the order of the $\cmin$ and $\cmax$ segments, its expressions can be obtained by swapping $\alpha\leftrightarrow\beta$ in those for its maximally soft counterpart.
The $n_B(\mu_B)$ function for the stiff EOS is found to be
\begin{equation}\label{eq:nmustiff}
n_B(\mu_B)=
\begin{cases}
n_\tov\lrp{\frac{\mu_B}{\mu_\tov}}^\alpha, \quad &\mu_\tov \leq\mu_B\leq \mu_2\\
n_\pqcd\lrp{\frac{\mu_B}{\mu_\pqcd}}^\beta, \quad &\mu_2 \leq\mu_B\leq \mu_\pqcd
\end{cases}
\end{equation} 
where the transition point $(\mu_2, n_2)$ is given by
\begin{align}
	\mu_2&=\mu_\pqcd\lrp{\frac{\mu_\pqcd}{\mu_\tov}}^{\frac{\delta}{1-\delta}} \lrp{\frac{n_\tov}{n_\pqcd}}^{\frac{1}{\beta-\alpha}},\\
	n_2&=n_\pqcd \lrp{\frac{n_\tov}{n_\pqcd}}^{\frac{1}{1-\delta}} \lrp{\frac{\mu_\pqcd}{\mu_\tov}}^{\frac{\alpha}{1-\delta}},
\end{align}
and is marked by the blue ``+'' in \cref{fig:demo}.
It follows that
\begin{multline}
\Delta P_\mathrm{min}=\int^{\mu_2}_{\mu_\tov}\ud\mu\ n_\tov\lrp{\frac{\mu}{\mu_\tov}}^\alpha
\\
+
\int^{\mu_\pqcd}_{\mu_2}\ud\mu\ n_\pqcd\lrp{\frac{\mu}{\mu_\pqcd}}^\beta
\\
=\frac{n_\tov\mu_\tov}{\alpha+1}\lrb{\lrp{\frac{\mu_2}{\mu_\tov}}^{\alpha+1}-1}\\
	+\frac{n_\pqcd \mu_\pqcd}{\beta+1}\lrb{1-\lrp{\frac{\mu_2}{\mu_\pqcd}}^{\beta+1}}.
\end{multline}

Plugging in the expressions of $\mu_1$ and $\mu_2$,  $\Delta P_\mathrm{min}$ and $\dpmax$ simplify to
\begin{widetext}
\begin{align}
	\Delta P_\mathrm{max}&=\frac{n_\pqcd\mu_\pqcd}{\alpha+1}\lrb{1-\lrp{\frac{\mu_\tov}{\mu_\pqcd}}^{\frac{1+\alpha}{1-\delta} } \lrp{\frac{n_\pqcd}{n_\tov}}^{\frac{1+\alpha}{\beta-\alpha}} }
		+\frac{n_\pqcd\mu_\pqcd}{\beta+1}\lrb{\lrp{\frac{\mu_\tov}{\mu_\pqcd}}^{\frac{1+\alpha}{1-\delta} } \lrp{\frac{n_\pqcd}{n_\tov}}^{\frac{1+\alpha}{\beta-\alpha}}-\frac{n_\tov\mu_\tov}{n_\pqcd\mu_\pqcd}}, \label{eq:dpmax2} \\
	\Delta P_\mathrm{min}&=\frac{n_\tov\mu_\tov}{\alpha+1}\lrb{
	\lrp{\frac{\mu_\pqcd}{\mu_\tov}}^{\frac{1+\alpha}{1-\delta} } \lrp{\frac{n_\tov}{n_\pqcd}}^{\frac{1+\alpha}{\beta-\alpha}}
	-1}
	+\frac{n_\tov\mu_\tov}{\beta+1}\lrb{\frac{n_\pqcd \mu_\pqcd}{n_\tov\mu_\tov}-\lrp{\frac{\mu_\pqcd}{\mu_\tov}}^{\frac{1+\alpha}{1-\delta} } \lrp{\frac{n_\tov}{n_\pqcd}}^{\frac{1+\alpha}{\beta-\alpha}}}.\label{eq:dpmin2} 
\end{align}
\end{widetext}

Next we discuss the procedure to find $\maxcmin$,
the highest $\cmin$ that satisfies the inequality \cref{eq:dpmaxbound}, or equivalently the minimum of $\beta=1/\cmin$.
It saturates the $\dpmax$ bound \cref{eq:dpmaxbound} and is implicitly given by
\begin{equation}\label{eq:solve_maxcmin}
Z=\frac{1}{1+\alpha}\lrb{1-A}+\frac{1}{1+\beta}\lrb{A-\frac{x}{y}},
\end{equation} 
where for notational simplicity we have defined dimensionless quantities $x=\frac{\mu_L}{\mu_H}$, $y=\frac{n_H}{n_L}$, $A=x^\frac{1+\alpha}{1-\delta}y^\frac{1+\alpha}{\beta(1-\delta)}$, and $Z=\frac{P_H-P_L}{\mu_Hn_H}$.
This is a transcendental equation in $\alpha$ and $\beta$ and to the best of the author's knowledge no analytical solution exists.
We therefore solve $\min\lrc{\beta}\equiv\maxcmin$ numerically.

As discussed in the main text,  $\dpmax(\cmin)$ is monotonic.
As such, the root $\maxcmin$ if exists must be unique in and bracketed by the interval $[0, \cmax]$. 
Furthermore, the mean value bound \cref{eq:mvb} tightens the bracketing interval to $[0, \cmean]$. 
Thus, a simple bisection algorithm or methods of similar flavors is handy.\\

There are two scenarios in which \cref{eq:solve_maxcmin} does not have a solution.
In either circumstance, one can at least conclude $\cmin<\cmean$ (the mean-value bound), and it is also possible to get stronger bounds.
In the first scenario, the constraint \cref{eq:dpmaxbound} is not satisfied even if one takes $\cmin=0$.
This is the case for e.g. the APR EOS in \cref{fig:cmin_cmax1} when the pressure in cold quark matter is high $P/P_\mathrm{FD}\gtrsim0.5$. 
Since $\cmin=0$ yields the highest $\dpmax$, this violation indicates inconsistencies among the NS EOS, the value of $\cmax$, and pQCD predictions at $\mu_\pqcd$.
Entrusting the pQCD EOS and bound on $\cmax$ (e.g. $\cmax=1$)
one can thus rule out the NS EOS under consideration~\cref{eq:dpmaxbound} \cite{Komoltsev:2021jzg,Gorda:2022jvk,Somasundaram:2022ztm,Zhou:2023zrm}.
However, NS EOSs excluded this way can be made compatible with \cref{eq:dpmaxbound}.
The minimal alteration one can implement is to insert a FOPT at the onset density $n_\mathrm{crit}<n_\mathrm{TOV}$ where 
violation of \cref{eq:dpmaxbound} begins.
Note that this does not imply FOPT inside NSs, because the required FOPT is too strong 
to accommodate stable NSs beyond $n_\mathrm{crit}$.
The end results are 
(1) moderately reduced $n_\tov^\mathrm{new}=n_\mathrm{crit}$;
(2) at most a few percent reduction to $M_\tov$ (so static NS observables are barely affected); and 
(3) strong FOPT at densities higher than those reached in NSs. 
For details about this modification and the FOPT, see the preceding paper ref~\cite{Zhou:2023zrm}.

Alternatively, \cref{eq:dpmax} does not admit a solution when \cref{eq:dpmaxbound} holds even for $\cmin=\cmean$.
From \cref{fig:demo}, one finds that when $\cmin=\cmean$ the maximally soft EOS reduces to a single segment defined by
\begin{equation}\label{eq:meanEOS}
    C_s(\mu_B)=\cmean,\quad \mu_\tov\leq\mu_B\leq\mu_\pqcd.
\end{equation}
Further increasing $\cmin$ would turn the maximally soft EOS (red) 
into the maximally stiff model (blue). 
Therefore, in this scenario the upper limit $\maxcmin$ comes from the requirement 
\begin{equation}\label{eq:dpminbound}
    P_\pqcd-P_\tov\geq\dpmin(\cmin)
\end{equation}
due to the maximally stiff EOS, and \cref{eq:dpmaxbound} is unconstraining.
This happens if $\Delta P<\Delta P_\mathrm{mean}$, where
\begin{equation}\label{eq:dpmean}
	\Delta P_\mathrm{mean}=\frac{n_\pqcd \mu_\pqcd}{\cmean^{-1}+1}\lrb{1-\lrp{\frac{\mu_\tov}{\mu_\pqcd}}^{\cmean^{-1}+1}}
\end{equation} 
is the increment in pressure from $\mu_\tov$ to $\mu_\pqcd$ associated with the mean-value EOS~\cref{eq:meanEOS}.

In the main text, the bound \cref{eq:dpminbound} is ignored as its effect is sensitive to the strength of non-perturbative contributions to the cold quark matter EOS~\cite{Zhou:2023zrm}.
This choice renders our results conservative, since in the absence of \cref{eq:dpminbound} $\cmin$ is only constrained by the mean value bound \cref{eq:mvb} when $\Delta P<\dpmean$.
Including \cref{eq:dpminbound} generally yields stronger constraints on $\cmin$, though the exact value depends on the size of the superconducting pairing gap.

Unlike the circumstances for the mean-value bound or for the $\dpmax$ bound where ignoring non-perturbative effects leads to conservative constraints, setting superconducting gap to zero in quark matter would {\em underestimate} $\maxcmin$ from \cref{eq:dpminbound}.
This is because $\dpmin(\cmin)$ is monotonic increasing (see \cref{eq:nmustiff} or \cref{fig:demo}), so that an underestimated left-hand side of \cref{eq:dpminbound} leads to an lower-than-expected $\maxcmin$.

\newpage


%

\end{document}